\documentclass[a4paper,portrait]{article}
\usepackage[utf8]{inputenc}

\usepackage{amsfonts,amsmath,latexsym,amssymb}
\usepackage{authblk}
\usepackage[colorlinks=true,
allcolors=blue]{hyperref}
\usepackage{multirow}
\usepackage{multicol}
\usepackage{url}
\usepackage{color}
\usepackage{geometry}
\usepackage{graphicx}
\usepackage{subfigure}
\usepackage{stmaryrd}
\usepackage{caption}
\usepackage{indentfirst} 
\usepackage{tabularx, booktabs}
\usepackage{float}
\usepackage[table]{xcolor}
\usepackage[sorting=none, url=false, eprint=false, maxnames=5]{biblatex}
\usepackage{etoolbox} 
\patchcmd{\thebibliography}{\section*{\refname}}{}{}{} 

\usepackage{setspace}
\usepackage{notes2bib} 

\begin{document}

\title{Computational model for tumor response to adoptive cell transfer therapy}

\author[1]{L. M. Luque}
\affil[1]{Instituto de Física de Líquidos y Sistemas Biológicos - CONICET. La Plata, Argentina.}

\author[1,2]{C. M. Carlevaro}
\affil[2]{Departamento de Ingeniería Mecánica, Universidad Tecnológica Nacional, Facultad Regional La Plata, La Plata, Argentina.}

\author[3]{E. Rodríguez-Lomba}
\affil[3]{Hospital Universitario Gregorio Marañón. Madrid, España}

\author[4]{E. Lomba}
\affil[4]{Instituto de Química Física Rocasolano - CSIC. Madrid, España}

\date{}

\maketitle
\begin{abstract}

One of the barriers to the development of effective adoptive cell transfer therapies (ACT), specifically for genetically engineered T-cell receptors (TCRs), and chimeric antigen receptor (CAR) T-cells, is target antigen heterogeneity. It is thought that intratumor heterogeneity is one of the leading determinants of therapeutic resistance and treatment failure. While understanding antigen heterogeneity is important for effective therapeutics, a good therapy strategy could enhance the therapy efficiency. In this work we introduce an agent-based model to rationalize the outcomes of two types of ACT therapies over heterogeneous tumors: antigen specific ACT therapy and multi-antigen recognition ACT therapy. We found that one dose of antigen specific ACT therapy should be expected to reduce the tumor size as well as its growth rate, however it may not be enough to completely eliminate it. A second dose also reduced the tumor size as well as the tumor growth rate, but, due to the intratumor heterogeneity, it turned out to be less effective than the previous dose. Moreover, an interesting  emergent phenomenon results from the simulations, namely the formation of a shield-like structure of cells with low oncoprotein expression. This shield turns out to protect cells with high oncoprotein expression. On the other hand, our studies suggest that the earlier the multi-antigen recognition ACT therapy is applied, the more efficient it turns. In fact, it could completely eliminate the tumor. Based on our results, it is clear that a proper therapeutic strategy could enhance the therapies outcomes. In that direction, our computational approach provides a framework to model treatment combinations in different scenarios and explore the characteristics of successful and unsuccessful treatments.

\end{abstract}

\section{Author summary}
\label{sec:author}

In this work we analyze the outcomes of two types of adoptive cell transfer (ACT) therapies over heterogeneous tumors. On one hand we treat the tumor with one and two dosses of antigen specific ACT therapy. We found that one dose of antigen specific ACT therapy reduces the tumor size as well as its growth rate, however it is not enough to completely eliminate it. A second dose also reduces the tumor size and tumor growth rate, but, due to the intratumor heterogeneity, it is not as effective as the first dose. Moreover, we observe an  emergent phenomenon in our simulations, namely the formation of a shield-like structure that protects the most proliferative cells. On the other hand, we see that when applying a multi-antigen recognition ACT therapy, the tumor could be completely eliminated if the former is applied at an early stage of the tumor growth. Even though our model needs to be further extended to incorporate patient specific clinical data, these results are a promising step in the direction of a personalized tool for hypotheses testing and to facilitate and educated choice between different therapy protocols.

\section{Introduction}
\label{sec:introduction}

Adoptive cell transfer (ACT) therapy is a form of immunotherapy that is a rapidly growing area of clinical investigation which involves removing a patient’s or donor’s T-cells, growing and/or modifying them in a laboratory, and reinfusing them back to the patient \cite{act}.

There are currently three major modalities of ACT: tumor-infiltrating lymphocytes (TILs), genetically engineered T-cell receptors (TCRs), and chimeric antigen receptor (CAR) T-cells. TIL therapy involves expansion of a heterogeneous population of endogenous T-cells found in a harvested tumor, while CAR T-cells and TCRs involve expansion of a genetically engineered T-cell directed toward specific antigen targets. While successful application of ACT has been seen in hematologic malignancies \cite{act8, act9, act10}, its use in solid tumors is still in its early stages. One of the barriers to the development of effective cellular therapies, specifically for TCRs and CAR T-cells, is target antigen heterogeneity.

Intratumor heterogeneity (also known as intralesion heterogeneity) refers to distinct tumor cell populations with different molecular and phenotypic profiles within the same tumor specimen \cite{hetero2, hetero21}. It is associated with poor prognosis and outcome \cite{hetero23, hetero24, hetero25, hetero26}. It is thought that intratumor heterogeneity is one of the leading determinants of therapeutic resistance and treatment failure and one of the main reasons for poor overall survival in cancer patients with metastatic disease \cite{hetero21, hetero27}. Tumor heterogeneity has presented a considerable challenge to matching patients with the right treatment at the right time; therefore, it poses a challenge to accomplish the goals of precision medicine \cite{hetero28, hetero29}.

One strategy to overcome antigen escape and heterogeneity is through the use of a multi-antigen recognition circuit involving complementary antigens \cite{act77, act78}. One example of this is the syn-Notch receptor, which uses an engineered transmembrane receptor to induce expression of a tumor-specific CAR in response to recognition of an extracellular signal \cite{act78, act79}. However, since tumor cells share antigens with other non-cancerous cells in the human body, to target the antigen that is specific to tumor cells and avoid normal human tissue has been a crucial challenge for the development of cellular therapies. While strategies such as those based on syn-Notch receptors are promising, great care has to be taken to find therapy strategies that will both be effective and minimally toxic to the patient. Hence, the main goal of this work is to computationally model the response of a heterogeneous tumor to different strategies of ACT therapies. As it was mentioned before, intratumor heterogeneity has a large impact on the outcome of treatment and thus investigation into therapies strategies will help improve ACT therapies and select patients for whom the treatment is likely to be successful.

Within this broad context, mathematical and computational modeling have both contributed significantly to the understanding of how cancer develops and to outline different therapeutic strategies to improve patient outcomes. By predicting heterogeneous responses, they can help to reduce failures in clinical trials and establish effective drug regimens through computational virtual trials. A widely used modeling paradigm in the study of complex biological systems is the \textit{agent-based model} (ABM) \cite{ABM2, ABM1}. ABM are implemented mainly to simulate the actions, behaviors and interactions of autonomous individual or collective entities, with the aim of exploring the impact of an agent or a type of behavior in the system.

An agent is the smallest unit in this model, and it can exhibit different types of stochastic behavior, including interaction with other agents. Although these models simplify many aspects of reality, they have been shown to be extremely useful in a wide number of circumstances \cite{ns10, ns11, ns12}. In cancer research, these models are emerging as valuable tools to study emergent behavior in complex ecosystems \cite{ns13}, and are used to study the mutational landscape of solid tumors \cite{ns16, ns17}. Furthermore, they are increasingly used to optimize therapies, for example radiation therapy of solid tumors \cite{ns18}. Also, some models of immune-cell interactions have been proposed \cite{nortonreview}. Although these studies gave important insight into parts of the tumor-immune interaction, they did not investigate therapeutic strategies. By adjusting model parameters and simulation rules, the characteristics of successful and unsuccessful treatments can be explored to learn how therapy outcomes vary with a patient's tumor characteristics \cite{macklin16, macklin17, macklin18}. Cancer immunotherapy could thus benefit from simultaneously employing molecular approaches (what medicinal chemistry can be employed to target specific molecular biology?) and multicellular systems-level approaches (what therapy protocol will lead to the best cancer control and induce remission?).

This work introduces a computational multiscale agent-based model to study immunosurveillance against heterogeneous tumors, with a special focus on the spatial dynamics of stochastic tumor–immune contact interactions. It could predict tumor response to different therapeutic strategies in order to discern whether a tumor is likely to respond to treatment or not. The model can be adjusted to reflect specific types of cancer to enable quantitative predictions of therapy–biomarker combinations and to be used as a platform for conducting virtual clinical trials.

The manuscript is organized as follows: After detailing the agent-based model in section \ref{sec:methods}, results are presented in section \ref{sec:results}. Discussion and future directions are found in Section \ref{sec:discussion}.

\section{Materials and methods}
\label{sec:methods}

The model presented herein  builds upon  previous work by Luque et al. on tissue growth kinetics \bibnote[mipaper]{L. M. Luque, C. M. Carlevaro, C. J. Llamoza Torres, E. Lomba; ``Physics-based tissue simulator to model multicellular systems: A study of liver regeneration and hepatocellular carcinoma recurrence''}. The following subsections will briefly recall details of the mentioned  model. Subsequently in subsections \ref{sec:methods;subsubsec:hetero} and \ref{sec:methods;subsubsec:immuno}, we will comment on the  the new features related to intratumoral heterogeneity and immunosurviellance modules implemented in this work.

\subsection{Model setup}

Our model is implemented resorting to an object oriented programming model, and to that aim C++11 language have been used. Simulation CPU time depends on model parameters such as domain (lattice) size, cell number and simulation length (in time); a typical simulation run takes approximately $6$ h on a single core of an Intel i7-10510U CPU. Model visualization is performed with Ovito \cite{ovito}, Paraview \cite{paraview} and Matplotlib \cite{matplotlib}.

\subsection{Diffusion solver}
\label{sec:methods;subsec:diffusion}

Cell behaviour is mostly dependent on the values and gradients of diffusing substrates in the tumor microenvironment. Diffusion process is modeled as a vector of reaction-diffusion partial differential equations for a vector of chemical substrates. It is discretized over a Cartesian mesh for computational convenience, in such a way that each voxel (volumetric pixel) stores a vector of chemical substrates. Each substrate diffuses and decays, and can be secreted or uptaken by individual cells at their specific positions.

To model the effect of blood vessels, or to apply Dirichlet boundary conditions, the so-called Dirichlet nodes are also implemented. In that implementation, substrate values at any voxel within the simulation domain can be overwritten to turn the voxel into a continuous source of substrates.

\subsection{Cell agents}

In the context of cancer immunology, the agents represent cancer and immune cells. Their motion is governed by the balance of adhesive, repulsive, motile, and drag-like forces. It is important to note that  repulsive forces are really an elastic resistance to deformation.

One of the main features that makes our model different from others in the literature is that cells are off-lattice. Consequently, they are not confined to a particular lattice or spatial arrangement, they move through all space positions, and therefore underlying possible artifacts associated with the chosen lattice structure and spacing are removed.

Each cell has an independent cell cycle which is modeled as a directed graph, and can also progress through apoptotic and necrotic death processes. Any of the cell cycle (and death processes) time scales can be adjusted at the beginning of the simulation to match different types of growth and they can also be adjusted at any time on an individual cell in order to reflect the influence of its microenvironment.

As the cell progresses through its current cycle, it varies its volume (and sub volumes, such as nuclear volume, solid volume, fluid volume, etc.). These volumes are modeled with a system of ordinary differential equations that allow cells to grow or shrink towards a target volume.

\begin{figure}[!htb]
	\centering
	\includegraphics[width=0.9\linewidth]{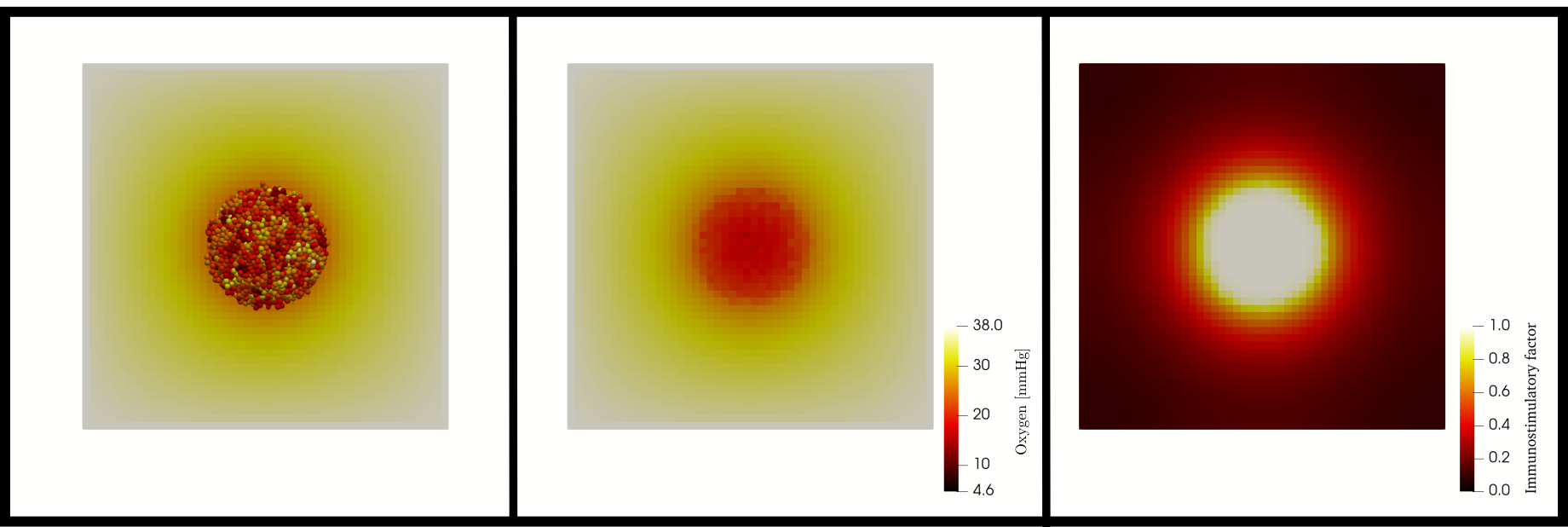}
	\caption{\textit{Substrate diffusion in the tumor microenvironment}. An heterogeneous tumor consuming oxygen (mmHg) for the micoenvironment, and secreting an immunoestimulatory factor (in arbitrary units).}
	\label{secrecion}
\end{figure}

As it was mentioned earlier, each cell can  secrete to or uptake from its chemical microenvironment, or sample the value or gradient of any or all substrates. This is very important since most of the cellular processes depend on the substrates that diffuse in the microenvironment. In every simulation step, each cell checks the substrate concentration in its voxel and base its behavior upon them. Figure \ref{secrecion} shows a tumor consuming oxygen from the microenvironment, and secreting an immunoestimulatory factor. This is one of the most important data structures of the cell because it links the cell with its microenvironment. Its inner workings are modeled by  a vector of partial differential equations which in practice implies the  addition of a cellular secretion/uptake term to the diffusion equation described in section \ref{sec:methods;subsec:diffusion}.

\subsection{Intratumor heterogeneity}
\label{sec:methods;subsubsec:hetero}

Even though eukaryotic cells replicate their DNA with astounding fidelity, the mechanism is not entirely error free. Every time a cell divides, a few mutational errors in the form of nucleotide substitutions and small deletions are introduced even in the absence of internal and external mutagens \cite{kike1, kike2}. Owing to the constant turnover of tumor cells and the large size of tumor cell populations, some of these stochastic mutational hits unavoidably affect genes with known cancer relevance, leading to the activation of oncogenes and/or inactivation of tumor suppressors, such as the p53 gene \bibnote[p53]{The TP53 gene provides instructions for making a protein called tumor protein p53 (or p53). This protein acts as a tumor suppressor, which means that it regulates cell division by keeping cells from growing and dividing (proliferating) too fast or in an uncontrolled way.}.

Among the many factors that drive tumor heterogeneity, genomic instability is most prominent in all malignancies. Many of the biological hallmarks associated with cancer development, such as limitless replicative potential, increase the mutational rate and genomic instability of malignant cells, which in turn give rise to other malignant traits \cite{hallmarks1, hallmarks2, hallmarks3}. This cascading effect often results in heterogeneity in the tumor as different cells acquire unique mutations that give rise to genetically distinct subpopulations \cite{kike3, kike4, kike5, kike6}. 

To study intratumor heterogeneity, each cancer cell is provided with a random expression of a mutant “oncoprotein”, $o$, using a normal distribution (a similar computational approach could be made to model intratumor heterogeneity based on the inactivation of the tumor suppressor p53 gene). This oncoprotein drives proliferation, \textit{i.e.} the greater the expression of $o$, the more likely the cell cycles and divides. In the absence of other selective pressures, the cells with the greatest $o$ expression clonally expand and dominate the dynamics of the tumor. Under the simplifying assumption that a highly-expressed mutant protein would be reflected as a more immunogenic peptide signature on major histocompatibility complexes (MHCs) \cite{macklin34}, each cell's immunogenicity is modeled as proportional to $o$.

\subsection{Immunosurviellance}
\label{sec:methods;subsubsec:immuno}

To model immunosurveillance T-cell agents are introduced. One of the main difference between T-cells and cancer cells present in our model, is that the former are  self-propelled. In other words, in addition to the forces due to the interaction with other cells and the basement membrane, immune cells move in response to chemical stimuli. As it was mentioned before, cancer cells secrete an immunostimulatory factor which diffuses through the microenvironment. Immune system cells perform a biased random migration towards this immunostimulatory gradient to find cancer cells. The migration performed along the direction \textbf{d}, which is updated according the immunostuimulatory factor gradient, is governed by the bias \textit{b}, which can take values $0 \leq b \leq 1$ where $0$ means Brownian motion and $1$ represents deterministic motion along \textbf{d}. Immune system cells change their migration velocity stochastically between $t$ and $t + \Delta t_{\text{mech}}$ with probability $\Delta t_{\text{mech}}/t_{\text{per}}$, where $t_{\text{per}}$ is the lymphocite’s mean persistence time. To change the velocity a random direction, $\mathbf{d}_r$,  is chosen by $\mathbf{d}_r = \left[ \sin{ \left( \phi \right) } \cos{ \left( \theta \right) }, \sin{ \left( \phi \right) } \sin{ \left( \theta \right), \cos{ \left( \phi \right) } }  \right] $, where $\theta$ is a random angle between $\left[0, \pi \right]$ and $\phi$ is a random angle between $\left[0, 2\pi \right]$. The migration velocity $\mathbf{v}_{mig}$ is then updated according to

\begin{equation}
    \mathbf{v}_{\text{mig}} = v_{\text{mot}}\frac{(1-b)\mathbf{d}_{r} - b\mathbf{d}}{|| (1-b)\mathbf{d}_{r} - b\mathbf{d} ||}
\end{equation}

\noindent where $v_{\text{mot}}$ is the migration speed. Notice that if the migration bias $b$ is $1$ the lymphocyte will perform a deterministic motion over the immunostimulatory factor gradient direction $\mathbf{d}$, while on the other hand, if $b=0$, it will perform a Brownian motion over the random direction $\mathbf{d}_{r}$.

If the immune cell is attached to a cancer cell, its velocity is set to zero. Finally, when updating the immune cell’s velocity, its migration velocity $\mathbf{v}_{\text{mig}}$ is added to the current velocity computed by the interaction with other cells.

T-cells continuously test for contact with cancer cells. In fact, if they detect contact, in any time interval, they have a probability of forming an adhesion regulated by $r_{\text{adh}} \Delta t$, where $r_{\text{adh}}$ is the rate of forming new adhesions. Once they form an adhesion they switch off their motility and cancer cells stop their cycling activity.

\begin{figure}[!htb]
	\centering
	\includegraphics[width=0.5\linewidth]{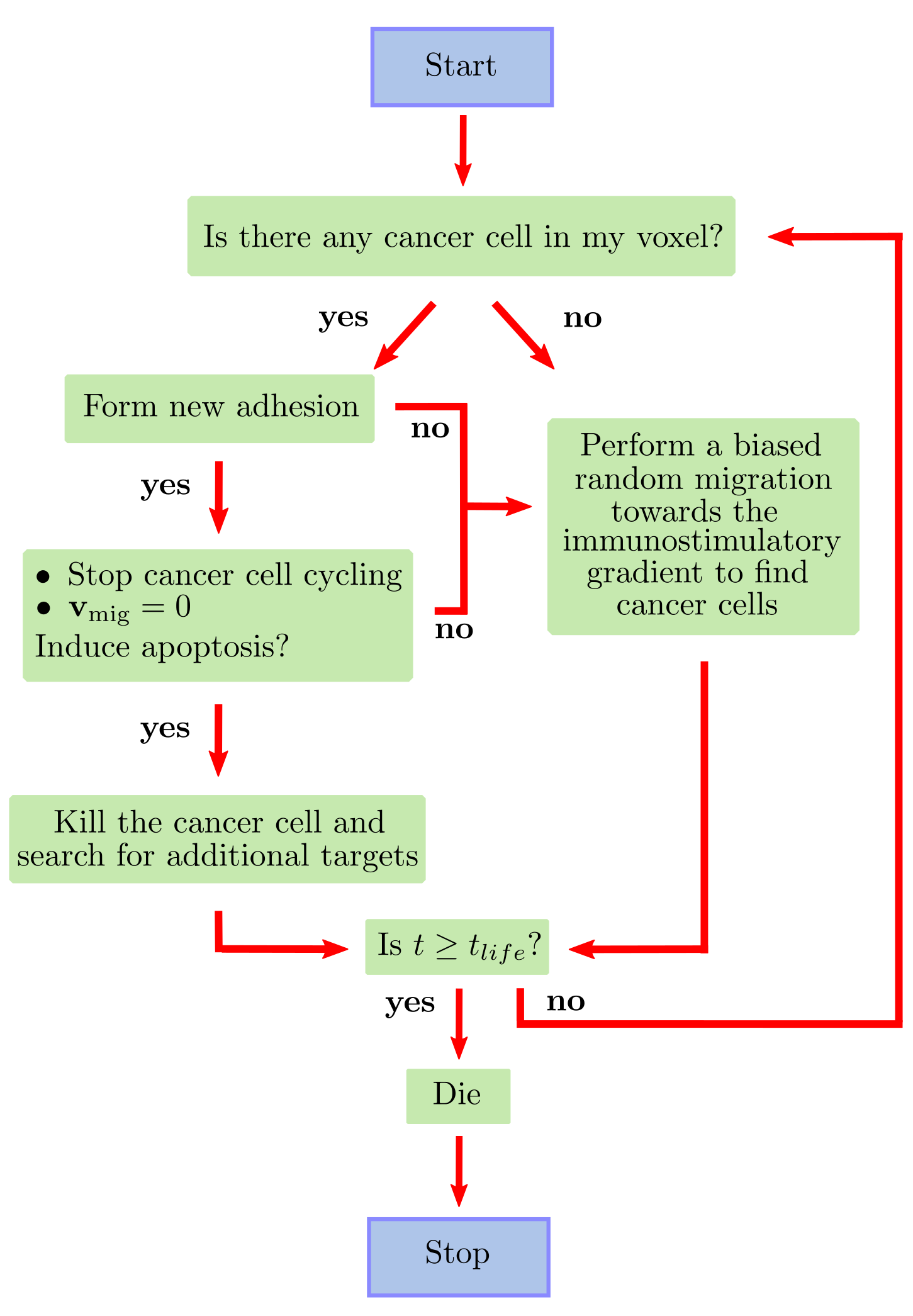}
	\caption{\textit{Immunosurviellance flow diagram.} $t_{life}$ represents the lifespan of the lymphocytes.}
	\label{diagrama}
\end{figure}

While adhered to a target cell, the immune cell agent attempts to induce apoptosis (e.g., by the FAS receptor pathway \cite{macklin35}) with a probability that scales linearly with immunogenicity. If successful, the tumor cell undergoes apoptosis, while the immune agent detaches and resumes its chemotactic search for additional tumor cell targets. If the immune cell does not kill the tumor cell, it remains attached while making further attempts to induce apoptosis until either succeeding or reaching a maximum attachment lifetime, after which it detaches without inducing apoptosis. In our model, T-cells have a lifespan of $10$ days and do not proliferate. A schematic representation of the inner workings of lymphocytes is depicted in Fig. \ref{diagrama}

\section{Results}
\label{sec:results}

\subsection{Heterogeneous tumor response to antigen specific ACT therapy}

Simulations take place in a grid of size $1000 \times 1000 \times 1000$ $\mu\text{m}$. A spherical tumor of $3963$ cells was seeded at the center of the simulation box. Each cell is assigned a mutant oncoprotein using a normal distribution that goes from $0$ to $2$ with a mean equal to $1$ and a standard deviation of $0.25$. For practical reasons, cells are labeled to reflect their oncoprotein expression: Type 1 ($1.5 \leq o < 2.0$), Type 2 ($1.0 \leq o < 1.5$), Type 3 ($0.5 \leq o < 1.0$), Type 4 ($0.0 \leq o < 0.5$). Cell proliferation and immunogenicity scale proportional to $o$, and an oncoprotein expression lower than $0.5$ is not enough to be recognized by T-cells.

\begin{figure}[!htb]
	\centering
	\includegraphics[width=\linewidth]{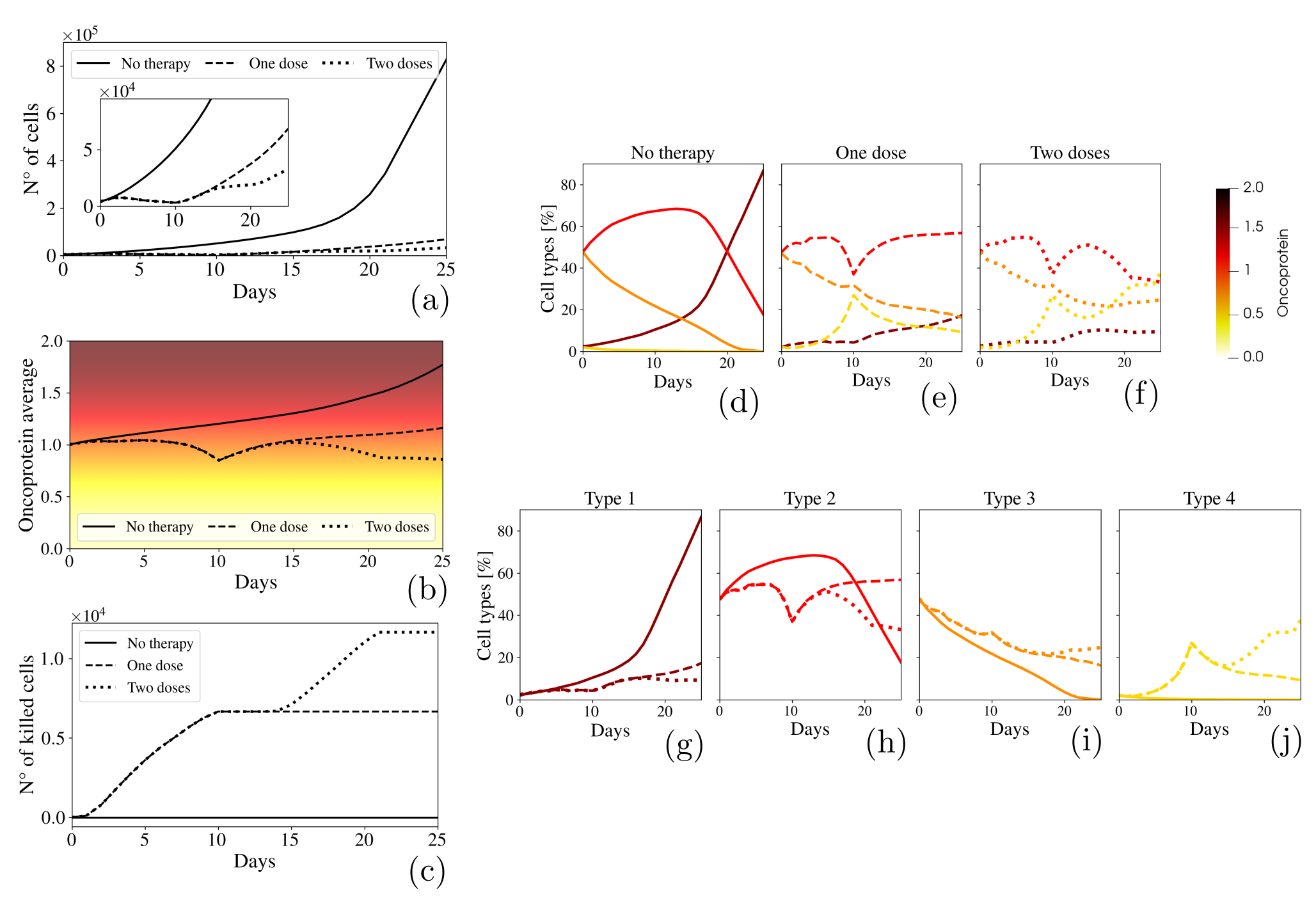}
	\caption{\textit{Heterogeneous tumor response to antigen specific ACT therapy.} The outcomes of a non treated tumor are represented by solid lines, while tumor responses after one and two doses of ACT therapy, are represented by dashed and dotted lines respectively. \textbf{(a)} Number of cancer cells. \textbf{(b)} Oncoprotein expression average. \textbf{(c)} Number of cancer cells killed by T-cells. \textbf{(d), (e), (f)} Percentage of the cell types that constitutes a non treated tumor as well as a tumor after one a two doses of ACT therapy. \textbf{(g), (h), (i), (j)} Comparison of the outcomes of non treated and treated tumors for the different cell types.}
	\label{antigen}
\end{figure}

As shown in figure \ref{antigen}a, without treatment the tumor grows fast due to the fact that the cells with higher oncoprotein expression, \textit{i.e.} the most proliferative cells, dominate its dynamics. It can be seen in figure \ref{antigen}d, which shows the percentage of cancer cell types inside the tumor, but is also reflected in the tumor's mean oncoprotein expression value (figure \ref{antigen}b). By the end of the simulation this value was between $1.5$ and $2.0$, that means that, despite the initial state of the tumor in which oncoprotein was normally distributed, it will evolve into a rapidly growing tumor.

Based on the scale of the simulated tumor, compared with those observed clinically, and considering the protocols reported in \cite{Linfocitos1, Linfocitos2, Linfocitos3} and references therein, $10000$ T-cells were randomly introduced at day $1$ to see how cancer evolution will change.
Figure \ref{antigen}a shows a drastically reduction in its growth rate, but not enough to completely eliminate it. Therefore, it is likely that the tumor will relapse. However, at the end of the simulation, the mean oncoprotein expression value shown in \ref{antigen}b, is considerably lower in comparison with the untreated tumor. These results suggest that even if one dose of ACT therapy is not enough to eliminate the tumor, it successfully decreases it size and reduces the rate of tumor growth. This can be explained by the fact that T-cells are more likely to kill the most proliferative cells of the tumor. It can be seen in figure \ref{antigen}e, which shows how type $1$ cells are no no longer dominating the tumor, as well as in figure \ref{antigen}g which shows a significant reduction in the percentage of type $1$ cells in comparison with a non treated tumor.

In order to test whether a complete elimination of the tumor was possible, a second dose of ACT therapy was applied at day $11$. As can be seen in figures \ref{antigen}a and \ref{antigen}b, a second dose also reduces the tumor size and the tumor growth rate, but it is not as effective as the first dose. This is reflected in the small differences found for the tumor size  mean oncoprotein values, but also in the number of cancer cells that T-cells were able to kill (figure \ref{antigen}c). This outcome results from the fact that the first dose eliminated most of type $1$ cells, which are the most likely to be killed by T-cells. Since immunogenicity scales proportionally to $o$, T-cells either do not recognize cancer cell due to the low oncoprotein expression, or spend more time sticking to targeted cells and trying to kill them (sometimes without success). This can be seen in figure \ref{antigen}f, in which a low percentage of type $1$ cells is present in the tumor, while the percentage of type $4$ cells (cancer ells that cannot be killed by T-cells) increased drastically from one dose to another (figure \ref{antigen}j).

\begin{figure}[H]
	\centering
	\includegraphics[width=0.9\linewidth]{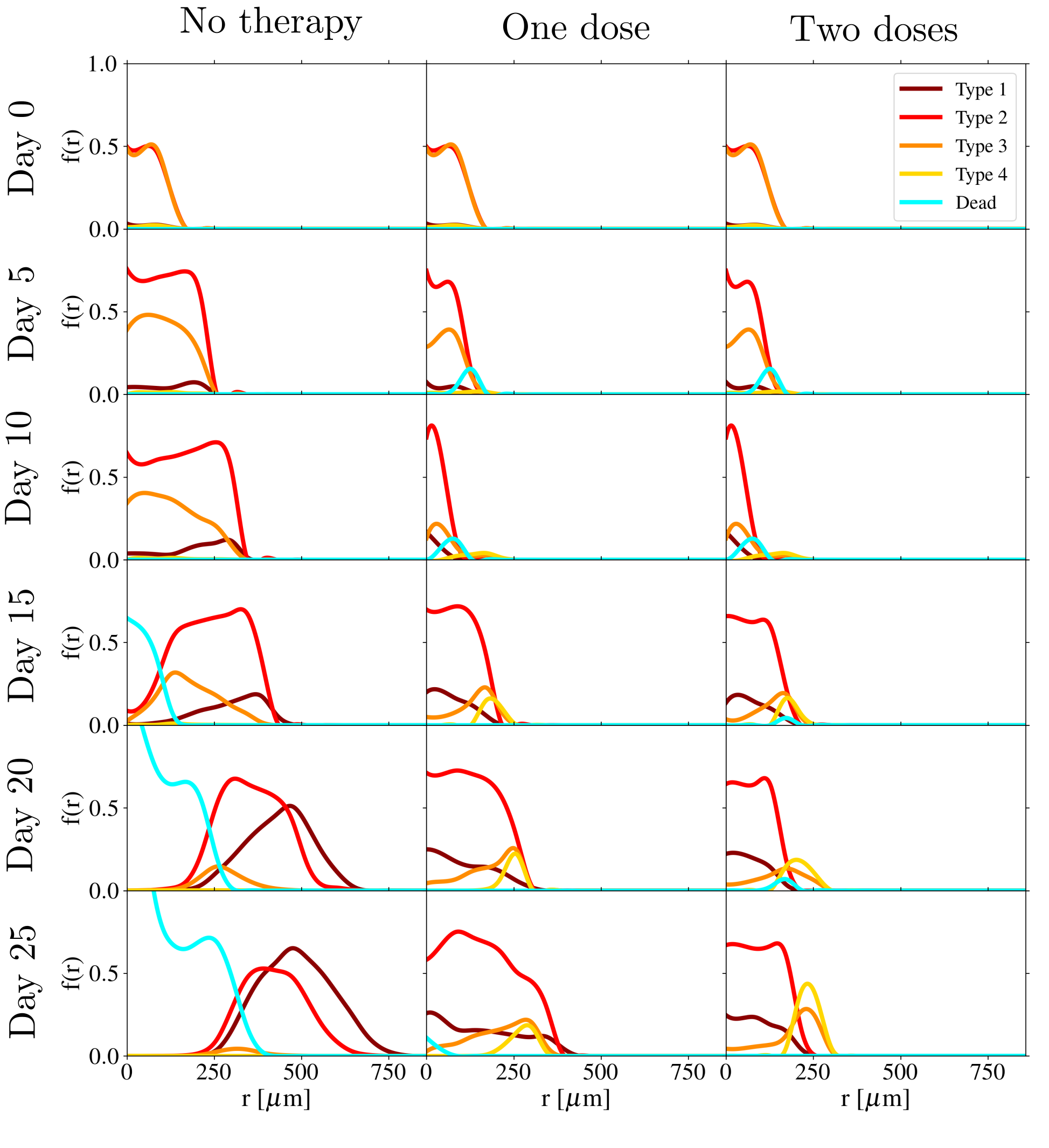}
	\caption{Radial distributions, $f(r)$, of different types of cells, in terms of the distance from the center of mass of each tumor to its surface. Left column shows a non-treated tumor, while the center column and the right column shows a tumor treated with one dose and two doses of ACT therapy respectively. Type $1$ cells are plotted in dark red, type $2$ in red, type $3$ in orange and type $4$ in yellow. Cyan curves represent dead cells, whether they have died for a T-cell attack or for lack of oxygen.}
	\label{distribucion}
\end{figure}

\begin{figure}[H]
	\centering
	\includegraphics[width=0.9\linewidth]{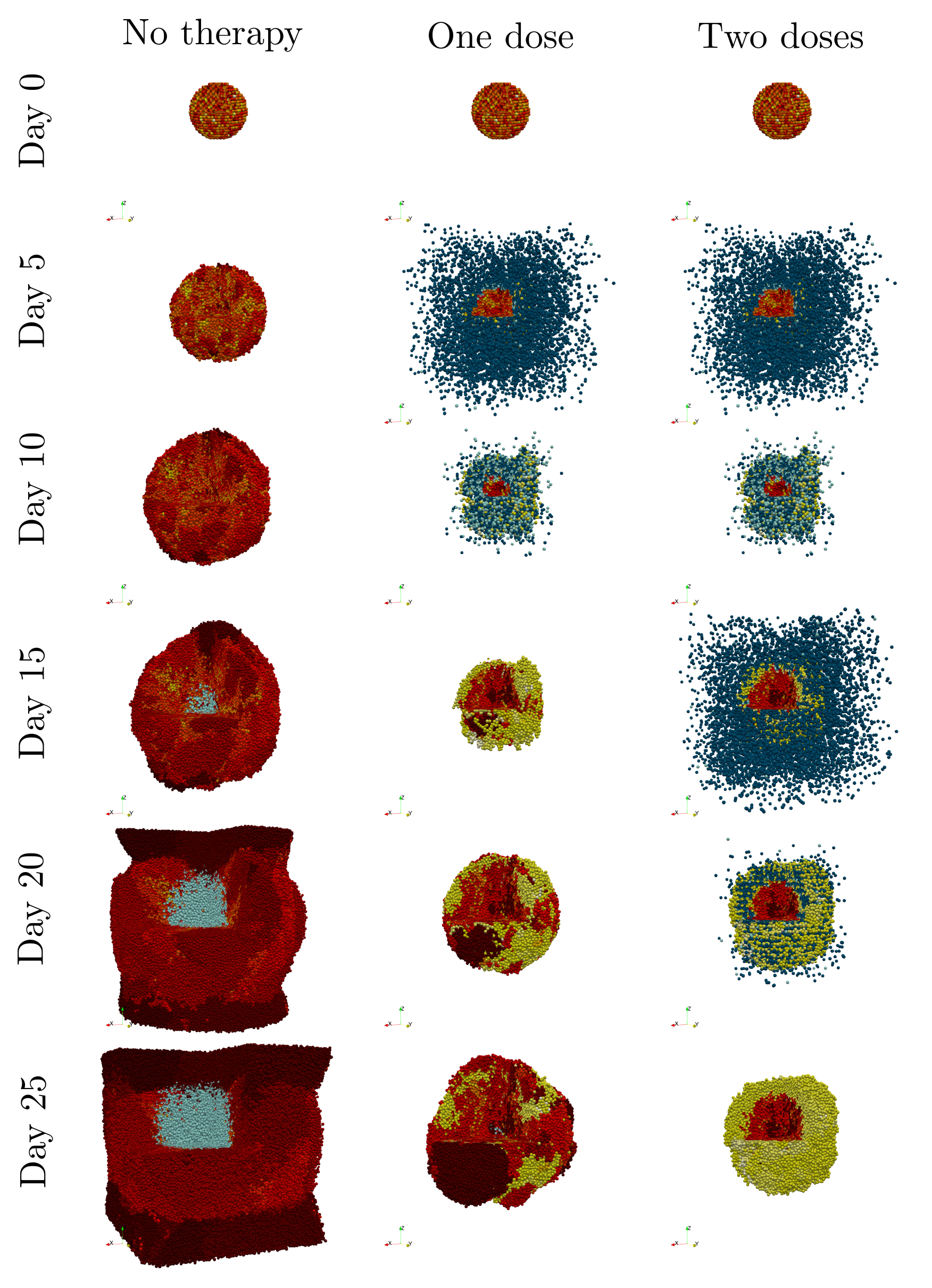}
	\caption{3D plot of the progression of a non treated tumor (left) tumor, and a tumor treated with one (center) and two (right) doses of ACT therapy, on specific days. T-cells are shown in dark blue, dead cells are shown in light blue. At day $25$, after two doses of ACT therapy a shield-like structure of cells with low oncoprotein expression is formed over cells with high oncoprotein expression. This leads to a reduction of therapy efficiency.}
	\label{tumores}
\end{figure}

Additionally, the use of a second dose of ACT gave rise to an interesting emergent phenomena. Type $4$ cells form a shield-like structure that prevents type $1$ and type $2$ cells to be reached by T-cells. To have a quantitative approximation of this behavior, figure \ref{distribucion} shows the radial distribution $f(r)$ of the different cell types inside the tumor. The distance $r$ ranges from the center of mass of each tumor, to its surface and is divided in spherical shells of width, $\Delta r$, of about $5$ cells radii. Day $25$ shows clearly how type $1$ and type $2$ cells (\textit{i.e.} the more proliferative cells) take over the tumor dynamics, whereas after two doses of immunotherapy those cells substantially decrease in number and a shield of type $3$ and type $4$ cells forms around them. This might be one of the reasons why a second ACT dose looses efficiency. These processes are qualitatively illustrated in figure \ref{tumores}.

Animations of the heterogeneous tumor response to one and two dosses of antigen specific ACT therapy can be seen in the Supplementary Material S1 Video.

\subsection{Heterogeneous tumor response to multi-antigen recognition ACT therapy}

Based on the previous results, a multi-antigen recognition type of therapy, such as syn-Notch receptor, was considered. In this approximation, T-cells can target every cancer cell, regardless of its oncoprotein expression value. Therefore different therapy strategies were tested.

\begin{figure}[!htb]
	\centering
	\includegraphics[width=\linewidth]{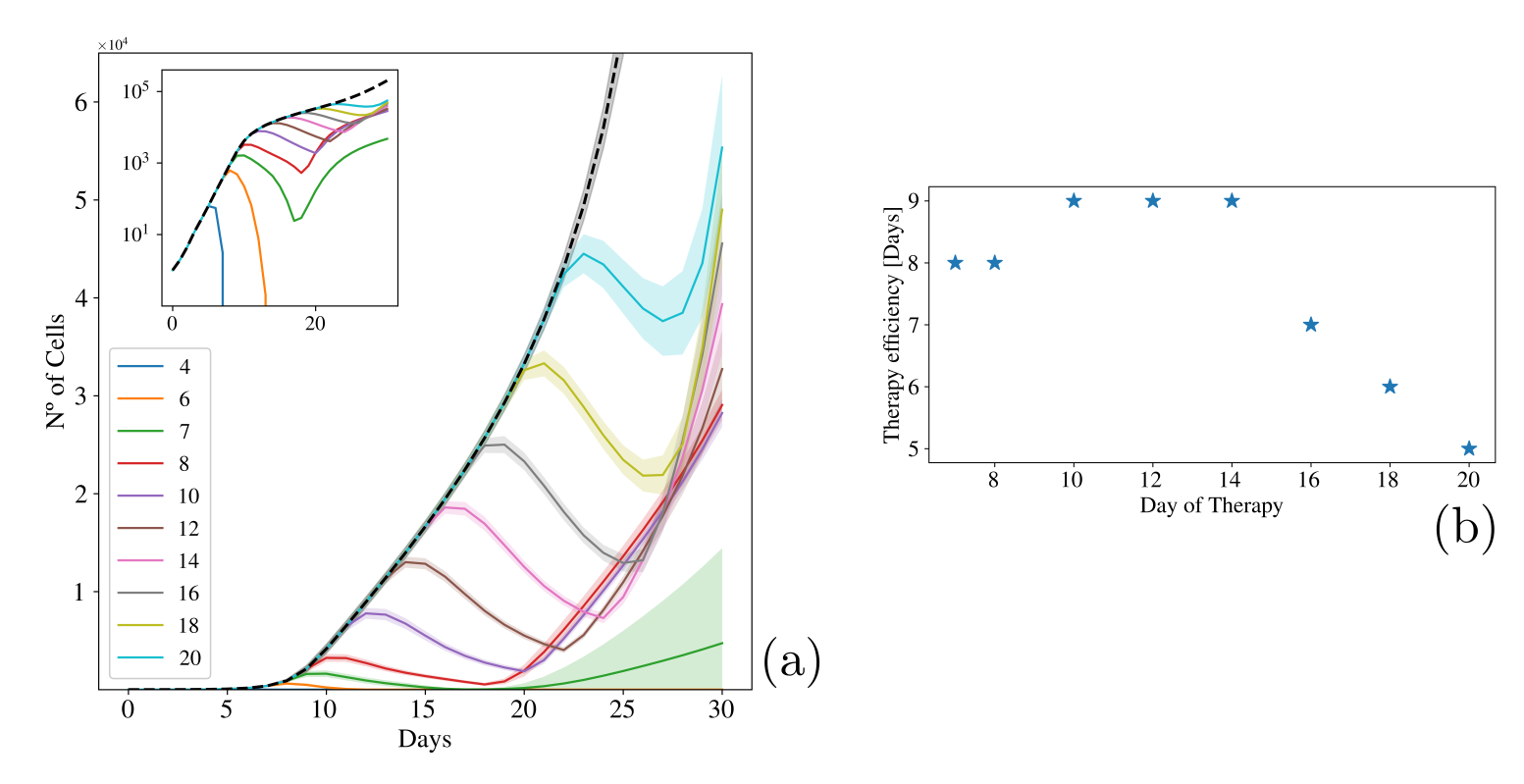}
	\caption{\textit{Heterogeneous tumor response to multi-antigen recognition ACT therapy} \textbf{(a)} Tumor response after one dose of ACT therapy applied at different days. Shaded regions represent the standard deviations of $20$ simulations. Inset shows the results in log scale. \textbf{(b)} Amount of days in which ACT therapy controls the tumor growth before its relapse.}
	\label{semillas}
\end{figure}

A single dose of $10000$ T-cells randomly introduced was applied at different stages of tumor growth. Figure \ref{semillas}a
shows the main results compared to a non treated tumor (black dashed line). 

Simulation suggests that when ACT therapy is applied at an early stage ($4$ and $6$ days after the beginning of the simulation), it will successfully eliminate the tumor. Since one of the milestones of this type of therapies is their capacity to target cancer cell not only in the primary tumor but in the whole body, this result is very promising if one consider this early stage tumor as an early stage metastasis.

After the $6^{th}$ day, ACT therapy shows a drastically reduction in tumor growth, but it is not enough to completely eliminate it. Moreover, as the application is delayed, the therapy loses its efficacy. On one hand, at an early stage, ACT therapy not only reduces the tumor size but it also decreases its growth rate, which can be deduced from the curve slope. However, by delaying the therapy application, this effect is not longer observed. On the other hand, figure \ref{semillas}b  shows that the time in which the therapy controls the tumor growth, \textit{i.e.} the therapy efficiency, also decreases when delaying the therapy application. Therefore, even though these therapies overcome antigen escape and heterogeneity, to determine an appropriate dosimetry strategy is necessary to produce effective therapeutic results.

\section{Discussion and future directions}
\label{sec:discussion}

One of the barriers to the development of effective cellular therapies, specifically for TCRs and CAR T-cells, is target antigen heterogeneity. It is thought that intratumor heterogeneity is one of the leading determinants of therapeutic resistance and treatment failure. While understanding antigen heterogeneity is important for effective therapeutics, a good therapy strategy could enhance the therapy efficiency.

Within this broad context, the aim of this work was to introduce an agent-based model that could rationalize the potential outcomes of ACT therapies over heterogeneous tumors using a computational approach.

When one dose of antigen specific ACT therapy is applied to a heterogeneous tumor, a drastically reduction in tumor size as well as in its growth rate is observed, however, it is not enough to completely eliminate it. Therefore, it is likely that the tumor will relapse. In order to test if a complete elimination of the tumor was possible, a second dose of therapy was applied. It also reduced the tumor size as well as the tumor growth rate, but it turned out to be less effective than the previous dose. Computational outcomes suggests that this lack of efficiency might be due to the fact that the first dose eliminated most of the high-oncoprotein expressing cells. Since immunogenicity scales proportional to the oncoprotein expression, $o$, T-cells either do not recognize low-oncoprotein expressing cancer cell, or spend more time trying to kill them (sometimes without success). One emergent phenomenon that came out of the simulations, and might be another reason for therapy inefficiency, was the formation of a shield-like structure of cells with low oncoprotein expression, that protected cells with high oncoprotein expression. While, to our knowledge, there is no reference to this type of shield-like barrier in ACT therapies, there are several works that study the physical barriers of the tumor cells that can reduce the therapy efficiency \cite{escudo}. Based on these results, one can predict the failure of a third dose of ACT therapy without targeting low-oncoprotein expressing cell first. Therefore, a different type of therapy or combination of therapies must be considered.

In order to overcome antigen escape and heterogeneity, another approach of ACT therapy, based in the syn-Notch receptor, has been studied. In this context T-cells can target every cancer cell, regardless of its oncoprotein expression value. It has been found that the earlier the therapy is applied, the more efficient it turns. In fact, it could completely eliminate the tumor. Since one of the milestones of this type of therapies is their capacity to target cancer cell not only in the primary tumor but in the whole body, this result is very promising if one consider this early stage tumor as an early stage metastasis. However, since cancers share antigens with other non-cancerous cells in the human body, great care has to be taken to find therapy strategies that will both be effective and minimally toxic to the patient.

There are several limitations of this model which point towards new directions for further development. One of the main constraints for its widespread use is the computational cost of the model. Even though thread parallelization in relevant sections of the algorithm is currently implemented, a full graphic processing units oriented re-writing of the most time consuming parts of the code is desirable. This will enhance the model's capacity to reach time-space scales that are unattainable so far. From a more practical standpoint, at this stage the model has  not been calibrated to any particular type of cancer. This an obvious handicap for its direct application in  the clinical practice. Clearly, a future line of work will have to focus on to tuning of model  parameters to  specific types of cancer. In this way, it will serve as a tool for hypotheses testing in the planning of alternative therapeutic protocols.

\section{Supporting information}
\label{sec:supporting}

\noindent\textbf{S1 Video. Heterogeneous tumor response to antigen specific ACT therapy.} Video available at \url{https://youtu.be/nyK98yZdQSs} 

\noindent\textbf{Source code.} The code used for running experiments is available at \url{https://github.com/lmluque/abm}

\section*{Acknowledgments}
\label{sec:acknowledgments}
This work was supported by the European Union Horizon 2020 Research
and Innovation Staff Exchange programme under the Marie Sk{\l}wodoska-Curie grant agreement No. 734276, which funded both the stays of E.L. at UNLP and M.C. and L.L. at IQFR-CSIC. E.L. also acknowledges funding from the Agencia Estatal de Investigaci\'on under grant no. PID2020-115722GB-C22.

\section{Author Contributions}
\label{sec:contributions}

\noindent\textbf{Conceptualization:} Luque, Carlevaro.

\noindent\textbf{Data curation:} Luque.

\noindent\textbf{Formal analysis:} Luque, Lomba, Carlevaro.

\noindent\textbf{Funding acquisition:} Lomba.

\noindent\textbf{Investigation:} Luque.

\noindent\textbf{Methodology:} Luque, Carlevaro.

\noindent\textbf{Project administration:} Lomba.

\noindent\textbf{Resources:} Lomba, Carlevaro.

\noindent\textbf{Software:} Luque.

\noindent\textbf{Supervision:} Lomba, Carlevaro.

\noindent\textbf{Medical supervision:} Rodríguez-Lomba.

\noindent\textbf{Validation:} Luque.

\noindent\textbf{Visualization:} Luque.

\noindent\textbf{Writing-Original draft preparation:} Luque.

\noindent\textbf{Writing-Review \& editing:} Lomba, Carlevaro, Rodríguez-Lomba.

\section{References}
\label{sec:references}
\printbibliography[heading=none]

\end{document}